\documentclass[12pt]{iopart}
\usepackage{iopams}

\expandafter\let\csname equation*\endcsname\relax
\expandafter\let\csname endequation*\endcsname\relax

\usepackage{amssymb}
\usepackage{amsmath}
\usepackage{soul}
\usepackage{cite}
\usepackage{enumerate}
\usepackage{empheq}
\usepackage{fancybox}
\usepackage{etoolbox}
\usepackage{hyperref}
\usepackage{color}
\makeatletter
\def\@mkboth#1#2{}
\newlength\appendixwidth
\preto\appendix{\addtocontents{toc}{\protect\patchl@section}}
\newcommand{\patchl@section}{%
  \settowidth{\appendixwidth}{\textbf{Appendix }}%
  \addtolength{\appendixwidth}{1.5em}%
  \patchcmd{\l@section}{1.5em}{\appendixwidth}{}{\ddt}%
}
\makeatother

\begin{document}

\title[Path integrals for fractional Brownian motion]{A unifying representation of path integrals for fractional Brownian motions}

\author{Olivier B\'enichou \& Gleb Oshanin}
\address{Sorbonne Universit\'e, CNRS, Laboratoire de Physique Th\'eorique de la Mati\`{e}re Condens\'ee (UMR CNRS 7600), 4 place Jussieu,
75252 Paris Cedex 05, France}

\date{\today}

\begin{abstract}
Fractional Brownian motion (fBm) is an experimentally-relevant, non-Markovian Gaussian
stochastic process with long-ranged 
correlations between the increments, parametrised  by the so-called Hurst exponent $H$;  
depending on its value the process can be sub-diffusive $(0 < H < 1/2)$, diffusive $(H = 1/2)$ or super-diffusive $(1/2 < H < 1)$. 
There exist three alternative equally often used  definitions of fBm --   
due to L\'evy and due to Mandelbrot and van Ness (MvN), which differ by the interval on 
which the time variable 
is formally defined.  
Respectively, the covariance functions of these fBms have different functional forms. Moreover, 
the MvN fBms have stationary increments, while for the L\'evy fBm this is not the case.
One may therefore be tempted to conclude that these are, in fact, different processes which only accidentally bear the same name.  
Recently determined explicit path integral representations also appear to have very different functional forms, 
which  
only reinforces the latter conclusion. 
Here
we develope
a unifying equivalent path integral representation of all three fBms in terms of Riemann-Liouville fractional integrals, 
which links the fBms  
and proves that they indeed belong to the same family. 
We show that the action in such a representation involves the fractional integral of the same form and order (dependent on whether $H < 1/2$ or $H > 1/2$) for all three cases, and differs only by the integration limits. 
\end{abstract}

\vspace{10pt}

\vspace{10pt}
\eads{olivier.benichou@sorbonne-universite.fr, gleb.oshanin@sorbonne-universite.fr}

\maketitle

\section{Introduction}

In the path integral formalism, one generalises the action principle of classical mechanics by replacing 
 a single classical "trajectory" by 
a functional integral over an ensemble of all possible trajectories. Such a formalism has been 
employed in the past for a theoretical analysis of diverse
physical  systems
\cite{hibbs,wiegel,kleinert,wio,leticia}, 
and also proven to be a powerful 
framework
for developing efficient simulation algorithms (see, e.g., \cite{ceperly}). 
Path integral representation of Brownian motion
permitted  to determine exactly statistical properties of several non-trivial
functionals of its trajectories  \cite{satya,dean,greg}, 
by taking advantage of 
the
Feynman-Kac theorem \cite{hibbs,kac}. 
The latter path integral representation
hinges on the
 Wiener's expression for the 
 probability   
 of observing on a time-interval $(0,T)$ a given realisation of a Brownian trajectory $x(t)$ \cite{wiener}:
 \begin{align}
 \label{prob}
 P[x(t)] \sim \exp\left(-S[x(t)]\right) \,,
 \end{align}
where $S[x(t)]$ is the action of the form 
\begin{align}
\label{action0}
S[x(t)] = \frac{1}{2} \int^T_0 dt \, \left(\frac{d x(t)}{dt}\right)^2 = \frac{1}{2} \int^T_0 dt  \,\xi_{\rm wn}^2(t) \,,
\end{align}
in which we have assumed, without lack of generality, that the diffusion coefficient is equal to $1/2$.  The dependence on the diffusion coefficient can be recovered 
by a mere rescaling of time.
In turn,  
$\xi_{\rm wn}(t)$ in eq. \eqref{action0} denotes  a given realisation of a zero-mean Gaussian white noise, 
which is linked to a Brownian trajectory $x(t)$ through the standard Langevin equation $d x(t)/dt = \xi_{\rm wn}(t)$, and has the covariance function 
\begin{align}
\label{noise}
\overline{\xi_{\rm wn}(t) \xi_{\rm wn}(t')} = \delta(t-t') \,. 
\end{align} 
with $\delta(t)$ being the Dirac delta-function. 
The bar in eq. \eqref{noise}  denotes  averaging with respect to different realisations
of noise.

 Note that in some physical situations the first-order derivative 
 in the quadratic action  \eqref{action0} is  
 naturally replaced by a higher-order derivative. Such a type of action arises, e.g., in analyses of conformations of stiff polymers or of membranes with bending rigidity \cite{kleinert,david}, or 
of the so-called random-acceleration process \cite{burk}. In all these examples the second derivative 
appears in place of the first one.
Extensions and generalisation of the action in eq. \eqref{action0}
have been obtained in case of coloured noises, noises entering in a non-linear form and some non-Markovian processes (see, e.g.,  \cite{wio}), as well as for some kinds of anomalous diffusions \cite{eule}, including the continuous-time random walks \cite{eule1} and fractional L\'evy motion \cite{1,2,3}.  The "optimal paths" $x_{\rm opt}(t)$, i.e., the action-minimising trajectories connecting fixed positions, say, the origin $0$ and a prescribed position $X_T$ achieved at time moment $T$,
 were determined in recent \cite{we0} for \textit{arbitrary} Gaussian stochastic processes; the derivation does not necessitate a knowledge of the explicit analogue of eq. \eqref{action0} for the process under study, and involves only  the covariance function ${\rm Cov}(t_1,t_2) = \overline{x(t_1) x(t_2)}$ of $x(t)$. It was shown in \cite{we0} that $S[x_{\rm opt}(t)] =  X_T^2/2 {\rm Cov}(T,T)$ and therefore $P \simeq \exp( - X_T^2/2 {\rm Cov}(t,t))$ provides then the short-time tail of the probability density function $\Psi_t$ of the first-passage time $t$ from $0$ to $X_T$ (see also \cite{Molchan,Krug,Guerin,Levernier1,Levernier2,Wiese} for other aspects of first-passage times of the FBM).

As it follows from the title of our paper, here we will be concerned specifically
 with fractional Brownian motion (fBm) - an experimentally-relevant (see, e.g., \cite{weiss,diego,vittoria}) 
 family of one-dimensional anomalous
stochastic processes with everlasting power-law correlations between the increments, for which no analogue of the
Feynman-Kac theorem exists. 
In fact, there are three alternative kinds of fBm, which are equally often considered in the literature
and
 differ between themselves by
 the time-interval $\Omega$ on which the time variable $t$ is defined, and eventually, by the form of the covariance function. 
The first one is due to
L\'evy \cite{levy}, who defined the process for a finite interval, 
  $t \in [0,T]$, as a fractional Riemann-Liouville integral (see eq. \eqref{RLintegral} below and \cite{kilbas}) of white noise.  
The second kind of fBm with $t \in (-\infty, \infty)$ was introduced by Kolmogorov \cite{kol}. 
Many properties of such a process and its relevance to diverse problems across the disciplines were later on discussed
by Mandelbrot and 
van Ness  \cite{mvn}. Nowadays,  this kind of fBm is commonly referred to as a two-sided Mandelbrot-van Ness fractional Brownian motion.
 Lastly, in the third case the time variable is defined on the positive half-axis, $t \in [0,\infty)$, and this process is called a one-sided Mandelbrot-van Ness fBm. 
 While for the two latter definitions the processes $x(t)$ have \textit{stationary} increments, for the L\'evy fBm the increments are not stationary. An extended overview  on applications of fBm can be found in recent \cite{ralf}.
 
We also note that in their analysis of the fBm  Mandelbrot and 
van Ness introduced an important notion of fractional Gaussian noise $\zeta_{\rm fGn}(t)$, as the noise generating 
a given fBm trajectory $x(t)$ via a Langevin-type equation, $\zeta_{\rm fGn}(t) = dx(t)/dt$. Clearly, this noise has very different statistical properties and covariance functions, as compared to the white noise $\xi_{\rm wn}(t)$. The statistical properties of such a noise also depend on the precise definition of the fBm.

Within the last three decades much progress has been achieved in 
determining the analogues of the action in eq. \eqref{action0} for all three kinds of fBm. 
The path integral representation has been first 
established for sub-diffusive Riemann-Liouville-type fBm in form of a fractional integral \cite{sebastian,sanchez} (see also \cite{wio}). 
For both one-sided and two-sided versions of the Mandelbrot-van Ness process, the early-stage analyses aimed at determining the action in form of a perturbation series expansion in powers of the parameter $\epsilon = H - 1/2$, where $H$ is the Hurst exponent (see below) with $H=1/2$ corresponding to the standard Brownian case with uncorrelated increments. Within this approach, the zeroth-order and several sub-leading terms in this series have been found \cite{wiese00,wiese0,wiese1,wiese2,wiese3}. Lastly, in recent \cite{we} explicit expressions for the action in eq. \eqref{action0} were derived for both one-sided and two-sided Mandelbrot-van Ness fBms with arbitrary values of the Hurst exponent.

The kernel functions in the analogues of eq. \eqref{action0} evaluated in \cite{we} are non-local, as compared to the one in eq. \eqref{action0}, and therefore embody essential correlations between the increments, which vanish only in the limit $H \to 1/2$.   
At the same time, the kernels obtained in \cite{sebastian,sanchez} and \cite{we} 
have very different functional forms and are also very different depending whether the processes are one-sided or two-sided. Given that the covariance functions of the L\'evy fBm
and of the Mandelbrot-van Ness ones are also very different, and that the increments are not stationary in the former case and stationary in the latter one, 
one may be tempted to conclude that the underlying processes  actually represent  quite \textit{different} and unrelated to each other stochastic processes, despite the fact that they happen accidentally to bear the same name. In this regard, it may be instructive to develop some alternative (as compared to the representations in \cite{sebastian,sanchez} and \cite{we}) unifying representation which 
will provide a clear evidence that all the above-mentioned fBm processes - L\'evy or Mandelbrot-van Ness ones, one-sided or two-sided, sub-diffusive and super-diffusive --  indeed are members of the same family.
In the present paper we show that such a natural unifying framework can be constructed by representing the kernel functions in terms of suitable fractional Riemann-Liouville integrals \cite{kilbas}, which have the same form and order, and differ by the integration limits only. 

This paper is outlined as follows. In Sec. \ref{def}
we 
introduce basic definitions, 
briefly discuss the properties of fractional Brownian motions and also recall available results for their explicit representations in terms of path integrals  (see \cite{sebastian,sanchez} and \cite{we}).  
 In Sec. \ref{main} we summarise our main results.  An interested reader can find a succinct outline of the derivations of our main results in Sec. \ref{calculations}, while  a more detailed account is presented in the Appendices.
 Finally, in Sec. \ref{conc} we conclude with a brief recapitulation of our results.

\section{Definitions and overview of previous results.}
\label{def}

As we have already remarked, fractional Brownian motion is a family of zero-mean Gaussian, non-Markovian processes with correlated increments, parametrised by the parameter $H$ - the Hurst index, which is a real number from the interval $(0,1)$, (bounded away from $0$ and from $1$). For $H \in (0,1/2)$, the fBm exhibits a sub-diffusive behaviour due to negative correlations between the increments, while for $H \in (1/2,1)$ the increments tend to have the same sign, which entails a super-diffusive motion.
 In the borderline case $H=1/2$, standard Brownian motion with independent increments is recovered. 
 
{\bf L\'evy fBm.} L\'evy \cite{levy} used the left-sided\footnote{Note that the conventional term "left-sided" (usually denoted by the symbol "+" put in the subscript),  simply means that the time variable $t$ appears on the upper terminal of integration, in contrast to the right-sided case (denoted by the symbol $"-"$) with the time variable appearing on the lower terminal. 
} Riemann-Liouville fractional integral of function $f(t)$ of the order $\alpha$ (see, e.g., \cite{kilbas}) :
\begin{align}
\label{RLintegral}
I^{\alpha}_{+,a}[f(t)] = \frac{1}{\Gamma(\alpha)} \int^t_a  \frac{ f(\tau)}{\left(t - \tau\right)^{1-\alpha}}  \, d\tau\,,
\end{align}
to define a given fBm trajectory $x(t)$ as
\begin{align}
\label{RL}
x(t) &= I^{H+1/2}_{+,0}[\xi_{\rm wn}(t)] = \frac{1}{\Gamma(H+1/2)} \int^t_0  \frac{\xi_{\rm wn}(\tau)}{\left(t - \tau\right)^{1/2-H}} \, d\tau \,, \quad x(t=0) = 0 \,,
\end{align}
where $\xi_{\rm wn}(t)$ is the above-defined Gaussian, delta-correlated white noise with zero mean value and the covariance function defined in eq. \eqref{noise}. For $H = 1/2$, the process in eq. \eqref{RL} is clearly the standard Brownian motion. 

The  covariance function   
${\rm Cov}(t_1, t_2)  = \overline{x(t_1) x(t_2)}$ of the fBm process in eq. \eqref{RL}
is given for $t_1 \geq t_2$ by
\begin{align}
\label{covRL}
{\rm Cov}(t_1, t_2)  &  = \frac{\left(H+1/2\right)}{\Gamma^2\left(H+3/2\right)}   \frac{t_2}{ (t_1 t_2)^{1/2-H}} \,_2F_1\left(1, \frac{1}{2}-H; \frac{3}{2}+H; \frac{t_2}{t_1}\right) \,,
\end{align}
while the expression for $t_2 \geq t_1$ is obtained from eq. \eqref{covRL} by merely interchanging $t_1$ and $t_2$. In eq.  \eqref{covRL} and henceforth, $\Gamma(z)$ and $_2F_1$ denote the Gamma function and the Gauss hypergeometric function, respectively. Note that the mean-square displacement of such a process obeys $\overline{x^2(t)} = t^{2H}/(2H \Gamma^2(H+1/2))$, as can be inferred  from eq. \eqref{covRL} directly, by setting $t_1=t_2=t$. Therefore, the process is sub-diffusive for $H < 1/2$, diffusive for $H = 1/2$ and super-diffusive  for $H > 1/2$.
In the limit $H \to 1/2$, the Gauss hypergeometric function $_2F_1$   in eq. \eqref{covRL} converges to $1$ and hence, the covariance function becomes ${\rm Cov}(t_1, t_2) = {\rm min}(t_1,t_2)$, as it should.

Path integral representation for the process in eq. \eqref{RL} has been found in \cite{sebastian,sanchez}  for a sub-diffusive (with $0<   H < 1/2$) Riemann-Liouville fBm. Using a simple argument based on the Gaussian nature of the underlying noise, it was shown (see the text after eq. (27)  in \cite{sebastian}) that the action $S[x(t)] $ can be written as 
\begin{equation}
\label{actionRL}
\begin{split}
S[x(t)] &= \frac{1}{2} \int^T_0 dt \left\{I^{1/2-H}_{+,0}\left[\frac{d x(\tau) }{d\tau}\right]  \right\}^2 \\&= \frac{1}{2} \int^T_0 dt \left\{\frac{1}{\Gamma(1/2-H)}  \int^t_0 \frac{d x(\tau) }{d\tau} \frac{d\tau}{(t - \tau)^{H+1/2}} \right\}^2 \,, \quad 0< H < 1/2 \,,
\end{split}
\end{equation}
where the expression in the curly brackets is, in fact, the reciprocal operator (the so-called fractional derivative) 
of the fractional integral in eq. \eqref{RL}. Note that we have somewhat changed the notations in eq. \eqref{actionRL} as compared to the ones used in \cite{sebastian}, to make them consistent with our subsequent analysis.

The representation \eqref{actionRL} is not valid for the super-diffusive L\'evy fBm. Indeed, for
$1/2 < H < 1$, the inner fractional integral in eq. \eqref{actionRL} is not defined, because the kernel has a non-integrable singularity. 
However,  a valid representation in the super-diffusive case can be obtained from eq. \eqref{actionRL} by a mere integration by parts. Assuming that the fractional Gaussian noise of the L\'evy-type (which generates a given trajectory $x(t)$) is zero for $t = 0$ and $t = T$, we find
\begin{equation}
\label{actionRLsup}
\begin{split}
S[x(t)] &= \frac{1}{2} \int^T_0 dt \left\{I^{3/2-H}_{+,0}\left[\frac{d^2 x(\tau) }{d\tau^2}\right]  \right\}^2 \\&= \frac{1}{2} \int^T_0 dt \left\{\frac{1}{\Gamma(3/2-H)}  \int^t_0 \frac{d^2 x(\tau) }{d\tau^2} \frac{d\tau}{(t - \tau)^{H-1/2}} \right\}^2 \,, \quad 1/2 < H < 1 \,.
\end{split}
\end{equation}
Note that in this representation the second-order derivative naturally replaces the first-order one, likewise it happens for a \textit{super-diffusive} random-acceleration process \cite{burk}. Note, as well,  that the singularity of the kernel is now integrable for $1/2 < H < 1$ and therefore, the expression \eqref{actionRLsup} is well-defined.  The derivation of the form in eq. \eqref{actionRLsup} using a canonical approach, which hinges on the integral representation of the Gauss hypergeometric function and application of the fractional integrals technique, will be presented elsewhere \cite{ol}.

{\bf Mandelbrot - van Ness fBm.} Mandelbrot and van Ness (MvN) \cite{mvn} 
used instead the Weyl integral \cite{kilbas} to define a given fBm trajectory $x(t)$ for $t \geq 0$ as 
\begin{equation}
\begin{split}
\label{MvN}
x(t) &= \frac{1}{\Gamma(H+1/2)} \Bigg\{  \int^t_0 (t-\tau)^{H-1/2} \xi_{\rm wn}(\tau) \, d\tau \\
&+ \int^0_{-\infty} \left[(t-\tau)^{H-1/2} 
- (-\tau)^{H-1/2}\right] \xi_{\rm wn}(\tau) \, d\tau 
\Bigg\} \,, \quad x(t=0) = 0 \,,
\end{split}
\end{equation}
and in a similar way for the evolution at $t < 0$, if one defines the time variable $t$ on the whole real line (two-sided process). Such a fBm process, which is also fixed to be at the origin at $t=0$,  
depends explicitly on the evolution of noise in the entire past. 
Respectively, the form of the covariance function depends 
on whether one considers the two-sided version of the process, with $t \in (-\infty,\infty)$, in which case
\begin{align}
\label{22}
{\rm Cov}(t_1, t_2) = \overline{x(t_1) x(t_2)} = \frac{1}{2} \left(|t_1|^{2H} + |t_2|^{2H} - |t_1 - t_2|^{2H}\right) \,,
\end{align}
or restricts $t$ to be only positive (one-sided process). In this latter case, one has
\begin{align}
\label{23}
{\rm Cov}(t_1, t_2) = \frac{1}{2} \left(t_1^{2H} + t_2^{2H} - |t_1 - t_2|^{2H}\right) \,.
\end{align}
Setting $t_1=t_2=t$ (with $t>0$) in eqs. \eqref{22} and \eqref{23}, one finds $\overline{x^2(t)} = t^{2H}$, which differs 
from the analogous result for L\'evy fBm given above only by a numerical factor.

For MvN fBms exact analogues of the action in eq. \eqref{action0} were determined in recent \cite{we}.  For \textit{sub-diffusive} one-sided MvN fBm 
 the action is given by
\begin{equation}
\begin{split}
\label{action1ssub}
S[x(t)] &=  \frac{{\rm ctg}(\pi H)}{4 \pi H} \int^{\infty}_0 \int^{\infty}_0  \frac{dx(t_1)}{dt_1}  \frac{dx(t_2)}{dt_2} \frac{{\rm I}_z(1/2-H,H)}{|t_1 - t_2|^{2H}}  dt_1 dt_2 \,,  \quad 0 < H < 1/2 \,, \\
&\qquad \qquad {\rm I}_z(a,b) = \frac{1}{B(a,b)} \int^z_0 x^{a-1} (1- x)^{b-1} dx \,, \quad z = \frac{4 t_1 t_2}{(t_1 + t_2)^2} \,,
\end{split}
\end{equation}
where $B(a,b)$ and ${\rm I}_z(a,b)$ are the beta- and the regularized incomplete beta-functions, respectively. 
In turn, for two-sided sub-diffusive fBm one has
\begin{equation}
\begin{split}
\label{action2ssub}
S[x(t)] &=  \frac{{\rm ctg}(\pi H)}{4 \pi H} \int^{\infty}_{-\infty} \int^{\infty}_{-\infty}  \frac{dx(t_1)}{dt_1}  \frac{dx(t_2)}{dt_2} \frac{dt_1 dt_2}{|t_1 - t_2|^{2H}}   \,, \quad 0 < H < 1/2 \,.
\end{split}
\end{equation}
For \textit{super-diffusive} MvN  fBm the action $S[x(t)]$ involves second-order derivatives with respect to the time variables. For the one-sided version of the fBm the action is given by
\begin{equation}
\begin{split}
\label{action1ssuper}
S[x(t)] &=  A \int^{\infty}_0  \int^{\infty}_0  \frac{d^2x(t_1)}{dt_1^2} \frac{d^2x(t_2)}{dt_2^2} \, |t_1 - t_2|^{2 - 2H} {\rm I}_{z'}(3/2-H,2H-2) \, dt_1 dt_2 \,, \\
& \qquad A =\frac{{\rm ctg}(\pi H)}{8 \pi H  (1-H) (2H-1)}  \,, \quad z' = \frac{{\rm min}(t_1,t_2)}{{\rm max}(t_1,t_2)} \,, \quad 1/2 < H < 1 \,,
\end{split}
\end{equation}
while in the two-sided case it obeys
\begin{equation}
\begin{split}
\label{action2ssuper}
S[x(t)] =  A \int^{\infty}_{-\infty}  \int^{\infty}_{-\infty}  \frac{d^2x(t_1)}{dt_1^2} \frac{d^2x(t_2)}{dt_2^2} \, |t_1 - t_2|^{2 - 2H}  \, dt_1 dt_2 \,, \quad 1/2 < H < 1 \,,
\end{split}
\end{equation}
where the numerical amplitude $A$ is defined in eq. \eqref{action1ssuper}.

 Several remarks are here in order:\\ a) Likewise eq. \eqref{action0} for the standard Brownian motion,  the 
 actions for the \textit{sub-diffusive} L\'evy fBm in eq. \eqref{actionRL} and for the \textit{sub-diffusive} MvN fBms in eqs.\eqref{action1ssub} and \eqref{action2ssub}, involve not the trajectories themselves, but rather the corresponding fractional Gaussian noises (or white noise in the case of standard Brownian motion). Together with eq. \eqref{prob} they define the probability of a given realization of a fractional Brownian noise. In turn, for the \textit{super-diffusive} case eqs. \eqref{actionRLsup}, \eqref{action1ssuper} and \eqref{action2ssuper}  involve the second-order derivatives of the trajectories and hence, the first-order derivatives of the respective fractional Gaussian noises. Recall that the action of the  above mentioned random-acceleration process also involves the second derivative of the trajectories \cite{burk}. \\
 b)  Despite the fact that both one-sided and two-sided MvN fBms have stationary increments, the kernels for the two-sided fBm depend only on the difference of the time variables, while for the one-sided fBm it is not the case, due to the incomplete beta-function which depends on both time variables.  \\
c)  A compact expression in eq. \eqref{action2ssub} with $H = 1/4$
 has been evaluated 
 about three decades ago in \cite{burl}, which aimed at finding an analogue of the Wiener's measure \eqref{prob} 
  for a given  
trajectory of a tagged monomer in a long Rouse polymer chain \cite{rouse,doi}. 
It was shown that for a chain which is 
initially in an equilibrium state, (i.e., the chain is left to evolve freely in a thermal bath
within an infinite period of time before the measurements start), 
 the action obeys precisely
the form in eq. \eqref{action2ssub} with $H =1/4$. This is not surprising, of course, given
 that the dynamics of a tagged bead in an initially pre-thermalised 
infinitely long Rouse chain corresponds to the two-sided MvN fBm process with $H=1/4$. \\
d) Expression  \eqref{action2ssub} with arbitrary $H \leq 1/2$ 
has been recently proposed in \cite{nech} as an effective Hamiltonian of topologically-stabilised polymers in melts, permitting to cover various conformations   
ranging from ideal Gaussian coils to crumpled globules. 
 
\section{Main results}
\label{main}
 
 We list here the main results of the present work for the MvN one- and two-sided fBms, relegating details of derivations to Sec. \ref{calculations} and the Appendices.  We show below that for the sub-diffusive MvN fBms the actions in eqs. \eqref{action1ssub} and \eqref{action2ssub} admit, respectively, the following representations
 in terms of the Riemann-Liouville fractional integrals :
\begin{equation}
\begin{split}
\label{main1a}
S[x(t)] &=  B \int^{\infty}_0 dt   \left\{ I^{1/2-H}_{-,\infty}\left[\frac{dx(t)}{dt}\right] \right\}^2 \\
&=  B \int^{\infty}_0 dt   \left\{\frac{1}{\Gamma(1/2-H)} \int^{\infty}_t  \frac{d x(\tau)}{d\tau} \frac{d\tau}{(\tau - t)^{H+1/2}}\right\}^2 \,, \quad 0 < H < 1/2 \,,
\end{split}
\end{equation}
and
\begin{equation}
\begin{split}
\label{main1b}
S[x(t)] &= B  \int^{\infty}_{-\infty}  dt \left\{I^{1/2-H}_{+,-\infty}\left[\frac{dx(t)}{dt}\right]\right\}^2 \\
&= B \int^{\infty}_{-\infty}  dt \left\{\frac{1}{\Gamma(1/2-H)} \int^t_{-\infty} \frac{dx(\tau)}{d\tau} \frac{d\tau}{(t - \tau)^{H+1/2}} \right\}^2 \,, \quad 0 < H < 1/2 \,,
\end{split}
\end{equation}
where in both representations the numerical amplitude $B$ is given by
\begin{equation}
\label{BBB}
B = \frac{1}{4 H \sin(\pi H) \Gamma(2H)} \,.
\end{equation}
Expressions \eqref{main1a} correspond to the one-sided case and involve the right-sided fractional Riemann-Liouville integral extended over the entire "future", while the expressions \eqref{main1b} hold for the two-sided case and involve the left-sided fractional integral 
extended over the entire "past".  Comparing the above expressions against the corresponding result for the L\'evy fBm, eq. \eqref{actionRL}, we notice that all three representations involve the fractional integral of the same order and differ only by the limits of integrations. Therefore, despite all the above-mentioned distinct features, these three kinds of sub-diffusive fBms can be considered as being the members of the same family.

For the super-diffusive MvN fBm we find
\begin{equation}
\begin{split}
\label{main2a}
S[x(t)] &= B \int^{\infty}_0 dt \left\{ I^{3/2-H}_{-,\infty} \left[\frac{d^2 x(t)}{dt^2}\right] \right\}^2 \\
&=  B \int^{\infty}_0 dt \left\{ \frac{1}{\Gamma(3/2-H)} \int^{\infty}_t  \frac{d^2 x(\tau)}{d\tau^2} \frac{d\tau}{(\tau - t)^{H-1/2}} \right\}^2 \,, \quad 1/2 < H < 1 \,,
\end{split}
\end{equation}
and
\begin{equation}
\begin{split}
\label{main2b}
S[x(t)] &=B \int^{\infty}_{-\infty} dt \left\{ I^{3/2-H}_{-,\infty} \left[\frac{d^2 x(t)}{dt^2}\right] \right\}^2 \\
&=  B \int^{\infty}_{-\infty} dt \left\{ \frac{1}{\Gamma(3/2-H)} \int^{\infty}_t  \frac{d^2 x(\tau)}{d\tau^2} \frac{d\tau}{(\tau - t)^{H-1/2}} \right\}^2 \,, \quad 1/2 < H < 1 \,,
\end{split}
\end{equation}
where the expressions \eqref{main2a} correspond to the one-sided, while the expressions \eqref{main2b}
-  to the two-sided super-diffusive MvN fBms. Note that the numerical amplitude $B$ (see eq. \eqref{BBB}) is the same for sub- and super-diffusive dynamics. 
Comparing the above expressions against the corresponding result for the L\'evy fBm, eq. \eqref{actionRLsup}, we realise that, again, all  
three representations involve the fractional integral of the same order and differ only in the integration limits. 
This signifies that also in the super-diffusive  case the three kinds of fBm evidently belong to the same family.

\section{Details of derivations}
\label{calculations}

We pursue here  the canonical approach which hinges on the following general expression for the action for an arbitrary  Gaussian  
process $x(t)$. This action is, in general, a quadratic functional 
of the form \cite{zinn},
\begin{align}
\label{action}
S[x(t)] &= \frac{1}{2}  \int_{\Omega} dt_1 \, x(t_1)   \int_{\Omega} dt_2 \, x(t_2) \, K(t_1,t_2) \,,
\end{align}
where $\Omega$ is the interval on which the time variable $t$ is defined. In turn, the kernel $K(t_1,t_2)$ in eq. \eqref{action} is a symmetric function of the time variables, i.e., $K(t_1,t_2) = K(t_2,t_1)$,  and is defined as the inverse 
of the covariance function of the process $x(t)$ :
\begin{align}
\label{int}
\int_{\Omega} dt_1 \, K(t_1,t_2) \, {\rm Cov}\left(t_1,t_3\right) = \delta\left(t_2 - t_3\right) \,.
\end{align}
Below we derive the solutions of the latter equation 
for the sub-diffusive one- and two-sided MvN fBms, as well as for their super-diffusive counterparts.
A brief account of the derivation of the kernels in the above-presented explicit form was given in \cite{we}. Here, we concentrate specifically on the solutions in the form of fractional integrals.

\subsection{Sub-diffusive Mandelbrot - van Ness fBms}

{\bf One-sided case.} 
In this case, the interval $\Omega$ on which the time variable is defined is $[0,\infty)$. Differentiating eq. \eqref{int} over the variable $t_3$, and taking advantage of the identities
\begin{equation}
\frac{d}{dt_3} \delta(t_2 - t_3) = - \frac{d}{dt_2}  \delta(t_2 - t_3) \,,
\end{equation}
and 
\begin{equation}
\frac{d}{dt_3} {\rm Cov}(t_1,t_3) = H \left(\frac{1}{t_3^{1 - 2 H}} - \frac{{\rm sign}(t_3-t_1)}{|t_3-t_1|^{1 - 2H}}\right) \,,
\end{equation}
we find that eq. \eqref{int}  becomes
\begin{align}
\label{BB}
\int_{0}^{\infty} dt_1 \, K(t_1,t_2) \, \frac{{{\rm sign}(t_3 - t_1)}}{|t_3 - t_1|^{1 - 2H}} = \frac{1}{H} \frac{d}{dt_2} \delta\left(t_2 - t_3\right) + \frac{1}{t_3^{1 - 2H}} \left(\int_{0}^{\infty} du \, K(u,t_2)\right) \,.
\end{align}
This is an integral equation with a Feller potential in the left-hand-side
and therefore, its formal solution is given by the inverse Feller transform (see, e.g., \cite{kilbas} and earlier paper \cite{lundgren}): 
\begin{align}
\label{B3}
 K(t_1,t_2)  &= \frac{{\rm ctg}(\pi H)}{2 \pi B(1/2-H,2H)} \frac{d}{dt_1} \int^{t_1}_0 \frac{du}{(t_1 - u)^{H+1/2}} \int^{\infty}_u \frac{dz}{(z-u)^{H+1/2}} \nonumber\\
 &\times \Bigg(\frac{1}{H} \frac{d}{d t_2} \delta(t_2 - z) + \frac{1}{z^{1-2H}}  \int_{0}^{\infty} du \, K(u,t_2)
 \Bigg) \,.
\end{align}
Performing the integrals in the right-hand-side of eq. \eqref{B3}, we realise that the contribution of the second term in the parentheses  vanishes, because the two-fold integral 
\begin{align}
\int^{t_1}_0 \frac{du}{(t_1 - u)^{H+1/2}} \int^{\infty}_u \frac{dz}{z^{1-2H} (z-u)^{H+1/2}} = \frac{4^H \pi^{3/2} \Gamma(1/2-H)}{\cos(\pi H) \Gamma(1-H)}
\end{align}
is equal to a  $t_1$-independent constant and hence, its derivative with respect to $t_1$ is zero.  The contribution of the first term in the parentheses can be conveniently represented as
\begin{align}
\label{BQ}
 K(t_1,t_2)  &= \frac{{\rm ctg}(\pi H)}{2 \pi H B(1/2-H,2H)} \frac{d^2}{dt_1 dt_2} \int^{\infty}_0 dt \frac{\theta(t_1 - t)  \theta(t_2 - t)}{[(t_1-t) (t_2-t)]^{H+1/2}}  \,,
\end{align}
where $\theta(z)$ is the Heaviside theta-function, such that $\theta(z) = 1$ for $z > 0$, and zero, otherwise.  Note that in this representation, the $t_1$- and  $t_2$-dependent contributions factorise. 
 Inserting the expression \eqref{BQ} into eq. \eqref{action} and changing the order of integrations, we get
 the following factorised representation
\begin{equation}
\begin{split}
S[x(t)]  = \frac{{\rm ctg}(\pi H)}{4 \pi H B(1/2-H,2H)} \int^{\infty}_0 & dt \left\{ \int^{\infty}_0 dt_1 \, x(t_1) \,  \frac{d}{dt_1} \frac{\theta(t_1 - t)}{(t_1 - t)^{H+1/2}} \right\} \\
&\times \left\{ \int^{\infty}_0 dt_2 \, x(t_2) \frac{d}{dt_2} \, \frac{\theta(t_2 - t)}{(t_2 - t)^{H+1/2}} \right\}  \,.
\end{split}
\end{equation}
Further on, performing the integration of each inner integral by parts, taking into account the initial condition $x(0)=0$  and also supposing that the trajectories $x(t)$ are not too over-stretched, (such that $\lim_{t \to \infty} x(t)/t^{H+1/2} \to 0$), we have
\begin{equation}
\begin{split}
\label{klm}
S[x(t)]  = \frac{1}{4 H \sin(\pi H) \Gamma(2H)} \int^{\infty}_0 dt \left\{ \frac{1}{\Gamma(1/2-H)} \int^{\infty}_0 d\tau \,  \frac{d x(\tau)}{d\tau} \,  \frac{\theta(\tau - t)}{(\tau - t)^{H+1/2}} \right\}^2 \,.  
\end{split}
\end{equation}
The last step consists in simply recalling the definition of the theta-function, which yields eventually our result in  eq. \eqref{main1a}.

The derivation of the action in eq. \eqref{action1ssub} from eq. \eqref{BQ}, and also the behaviour in the limit $H \to 1/2$, (in which case one recovers the standard Brownian motion result \eqref{action0}) is discussed in \ref{secA1}. 

{\bf Two-sided case.} 
In this case, the interval $\Omega$ on which the time variable is defined is the entire real axis, 
$t \in (-\infty,\infty)$. For the two-sided case, a link between the explicit form in eq. \eqref{main1b} and the form 
\eqref{action2ssub} involving the Riemann-Liouville integral is provided by the integral representation :
\begin{equation}
\begin{split}
\label{zuk}
&\frac{1}{|t_1 - t_2|^{2H}} = \frac{1}{B(1/2-H,2H)}  \int^{\infty}_{{\rm max}(t_1,t_2)} \frac{dt}{[(t - t_1) (t - t_2)]^{H+1/2}}  \\
&=\frac{1}{B(1/2-H,2H)}  \int^{\infty}_{-\infty} dt \frac{\theta(t - t_1) \theta(t-t_2)}{[(t - t_1) (t - t_2)]^{H+1/2}}  \,, \quad t_1,t_2 \in (-\infty,\infty) \,.
\end{split}
\end{equation}
Inserting this representation into eq. \eqref{action2ssub} and changing the integration order, we get
\begin{equation}
S[x(t)] = \frac{1}{4 H \sin(\pi H) \Gamma(2H)} \int^{\infty}_{-\infty} dt \left\{ \frac{1}{\Gamma(1/2-H)} \int^{\infty}_{-\infty} d\tau \,  \frac{d x(\tau)}{d\tau} \,  \frac{\theta(t - \tau)}{(t - \tau)^{H+1/2}} \right\}^2 \,,
\end{equation}
from which expression we obtain the result in  eq. \eqref{main1b}.    

Details of the derivation of eq. \eqref{action2ssub} can be found in \cite{we} and are also
 presented in a more complete form in  \ref{secA2}.

\subsection{Super-diffusive Mandelbrot - van Ness fBms}

{\bf  One-sided case.} For one-sided super-diffusive fBm it is convenient to take advantage of the following identity,
obeyed by the covariance function in eq. \eqref{23},
\begin{equation}
\label{rep}
\frac{1}{2} \left(t_1^{2H} + t_2^{2H} - |t_1 - t_2|^{2H}\right) = H(2H-1) \int^{t_1}_0 \int^{t_2}_0 \frac{du dv}{|u - v|^{2 - 2H}} \,.
\end{equation}
Inserting this identity into the integral equation \eqref{int}, and integrating by parts (see \ref{secA3} for more details), we find the following integral equation
\begin{equation}
\label{int3}
\int_0^{\infty} \frac{q(t_1,t_3) dt_1}{|t_1 - t_3|^{2 - 2H}} = \frac{\delta(t_3 - t_2)}{H (2H-1)} \,,
\end{equation}
obeyed by an auxiliary function $q(t_1,t_3)$ of the form
\begin{equation}
\label{quq}
q(t_1,t_3) = \int^{t_1}_0 d\tau_1 \int^{t_3}_0 d\tau_3 \, K(\tau_1,\tau_3) \,.
\end{equation}
The solution of eq. \eqref{int3} can be found in a standard way by applying the technique of fractional integrals (or can be found directly in \cite{lundgren}) and reads (see \ref{secA3})
\begin{equation}
\label{qus}
q(t_1,t_3) = \frac{1}{2 H \sin(\pi H) \Gamma(2 H) \Gamma^2(3/2-H)} \frac{d^2}{dt_1 dt_3} \int^{\infty}_0 dt \frac{\theta(t_1 - t) \theta(t_3 - t)}{[(t_1 - t) (t_3 - t)]^{H-1/2}} \,,
\end{equation}
such that the kernel $K(t_1,t_3)$ is given by
\begin{equation}
K(t_1,t_3) = \frac{1}{2 H \sin(\pi H) \Gamma(2 H) \Gamma^2(3/2-H)} \frac{d^4}{dt_1^2 dt_3^2} \int^{\infty}_0 dt \frac{\theta(t_1 - t) \theta(t_3 - t)}{[(t_1 - t) (t_3 - t)]^{H-1/2}} \,.
\end{equation}
Inserting the latter expression into eq. \eqref{action} and changing the integration order, we get
\begin{equation}
\begin{split}
\label{klmn}
S[x(t)] = \frac{1}{4 H \sin(\pi H) \Gamma(2H)} \int^{\infty}_0 dt \left\{\frac{1}{\Gamma(3/2 - H)} \int^{\infty}_0 d\tau \, x(\tau) \frac{d^2}{d\tau^2} \frac{\theta(\tau - t)}{(\tau - t)^{H - 1/2}}   \right\}^2 \,.
\end{split}
\end{equation}
Supposing next that the trajectories $x(t)$ are not too over-stretched at $t \to \infty$, (as compared to the typical behaviour $x(t) \simeq t^H$), 
such that
\begin{equation}
\lim_{t \to \infty}  \frac{x(t)}{t^{H + 1/2}} = 0 \,, \qquad \lim_{t \to \infty} \frac{1}{t^{H - 1/2}} \frac{d x(t)}{dt} = 0 \,,
\end{equation}
we find the representation of the action  in eq. \eqref{main2a} by integrating twice by parts the inner integral in eq. \eqref{klmn}. Convergence of the action  in eq. \eqref{main2a} to the standard result in eq. \eqref{action0} in the limit $H \to 1/2$ is demonstrated in \ref{secA3}.

{\bf  Two-sided case.} Following \cite{we}, here we pursue a bit different line of thought and first reformulate our eqs. \eqref{action} and \eqref{int} in terms of a two-sided fractional Gaussian noise $\zeta_{fGn}(t)$, rather than in terms of the fBm trajectories. These equations read
\begin{equation}
\label{S}
S\left[x(t)\right] = \frac{1}{2} \int_{-\infty}^{\infty}    \int_{-\infty}^{\infty}   \frac{d x(t_1)}{dt_1} \frac{d x(t_2)}{dt_2} \, q(t_1,t_2) \, dt_1 \, dt_2 \,, 
\end{equation}
where the kernel $q(t_1,t_2)$ is implicitly defined by the integral equation
\begin{equation}
\label{iop}
\int^{\infty}_{-\infty} dt_1 \, q(t_1,t_2) \, {\rm cov}(t_1, t_3) = \delta(t_2 - t_3) \,,
\end{equation}
with ${\rm cov}(t_1, t_3)$ being the covariance of the fractional Gaussian noise \cite{mvn}
\begin{equation}
\label{fGn}
{\rm cov}(t_1, t_3) = \overline{\frac{d x(t_1)}{dt_1} \frac{d x(t_3)}{dt_3}} = \frac{H (2H-1)}{|t_1 - t_3|^{2-2H}} \,. 
\end{equation}
Note that the kernel $q(t_1,t_2)$ is related to the kernel $K(t_1,t_2)$ in the original  eq. \eqref{action} 
through the integral transformation
\begin{equation}
q(t_1,t_2)  = \int^{t_1}_{-\infty} d\tau_1 \int^{t_2}_{-\infty} d\tau_2 \, K(\tau_1,\tau_2) \,,
\end{equation}
and therefore differs from a similar property defined in eq. \eqref{quq} only by the lower limit of integrations. 

An explicit form of the integral equation obeyed by 
$q(t_1,t_2)$ is obtained by simply inserting the definition in eq. \eqref{fGn} into the eq. \eqref{iop}, which gives
\begin{equation}
\label{equ}
\int^{\infty}_{-\infty} \frac{dt_1 \, q(t_1,t_2)}{|t_1 - t_3|^{2-2H}}  = \frac{1}{H (2H  - 1)} \delta(t_2 - t_3) \,.
\end{equation}
This is an integral equation with the Riesz potential in the left-hand-side and its solution is given by the inverse Riesz transform \cite{kilbas}. The solution presented in \cite{kilbas} yields directly the explicit representation in eq. \eqref{action2ssuper}, but is inconvenient for our purposes here. The point is that it contains a positive power of the difference $|t_1 - t_2|$ and therefore does not permit
 to get a factorised representation in terms of fractional integrals. To circumvent this difficulty,  we consider instead an integral equation for the auxiliary function 
\begin{equation}
u(t_1,t_2) = \int^{t_2}_{-\infty} d\tau_2 \, q(t_1,\tau_2) \,,
\end{equation}
and solve it in \ref{secA4} using the approach based on the fractional integrals. In doing so, we find that $u(t_1,t_2)$ obeys
\begin{equation}
\label{equ5}
u(t_1,t_2)  =  - \frac{{\rm ctg}(\pi H) \Gamma(H-1/2)}{2 \pi H \Gamma(3/2-H) \Gamma(2 H)} \frac{d}{dt_1} \int^{\infty}_{-\infty} dt \frac{\theta(t_1-t) \theta(t_2-t)}{[(t_1 - t) (t_2 - t)]^{H-1/2}}  \,,
\end{equation}
in which the dependence on $t_1$ and $t_2$ appears in a factorised form.

Further on, we concentrate on eq. \eqref{S} and integrate it by parts with respect to $t_2$. This gives, assuming that the fractional Gaussian noise vanishes at time equal to plus infinity,
\begin{equation}
\begin{split}
\label{lab}
S[x(t)] & = -  \frac{1}{2} \int^{\infty}_{-\infty} \int^{\infty}_{-\infty}  \frac{d x(t_1)}{dt_1} \frac{d^2 x(t_2)}{dt_2^2} \, u(t_1,t_2) \, dt_1 \, dt_2 \\
&= - \frac{1}{4 H \sin(\pi H) \Gamma(2H)} \int^{\infty}_{-\infty} dt \left\{ \frac{1}{\Gamma(3/2-H)} \int^{\infty}_t \frac{d^2 x(t_2)}{dt_2^2} \frac{dt_2}{(t_2 - t)^{H-1/2}}\right\} \\
&\qquad \qquad \qquad \times \left\{ \frac{1}{\Gamma(3/2-H)} \int^{\infty}_{-\infty} dt_1 \frac{d x(t_1)}{dt_1} \frac{d}{dt_1} \frac{\theta(t_1- t)}{(t_1 - t)^{H-1/2}}\right\} \,.
\end{split}
\end{equation}
Performing the integral in the last line of the latter equation by parts, we arrive at the desired representation in the last two lines of eq. \eqref{main2b}.  Convergence of the action in eq. \eqref{main2b} to the one in eq. \eqref{action0} in the limit $H \to 1/2$ is discussed in \ref{secA4}.

\section{Conclusions}
\label{conc}

To conclude, in the present paper we were concerned with the so-called fractional Brownian motions - non-Markovian Gaussian stochastic processes with long-ranged power-law correlations between the increments. In case when these correlations are positive prompting the increments between the successive moves to have the same sign, 
they entail a super-diffusive motion. On contrary, when correlations are negative such that the increments tend to have 
different signs, the resulting dynamical behaviour is sub-diffusive. 
Such processes, both super-diffusive and sub-diffusive, are often used in the literature 
to model various transport phenomena or systems with chemical reactions,  
and also efficient search processes, in which the underlying 
microscopic dynamics is anomalous. 
Concurrently, there is an ample evidence that many naturally-occurring processes 
are, in fact, fractional Brownian motions. 

In the literature, there exist three distinct definitions of fractional Brownian motions which differ basically by the interval on which the time variable is defined. These are the definition due to L\'evy, for which $t$ is defined on a finite interval, and two definitions 
due to Mandelbrot and van Ness, for which $t$ is defined either on the positive half-axis or on the entire real line, respectively. The corresponding covariance functions and the actions in the path-integral representations have also very different functional forms. Therefore, one may be prompted to conclude that these are, in fact, quite different and unrelated to each other classes of stochastic processes, and it is unclear why they
have the same common name. 

Motivated by this ambiguity, here we have developed a unifying framework which links all three processes together. 
Namely, we have established alternative (as compared to the explicit representations in \cite{we}) path integral representations of all three fractional Brownian motions in terms of Riemann-Liouville fractional integrals. 
We have shown that for all three definitions
 the action in such representations can be cast into the form which involves the fractional integral of the same kind and order (that depends only on whether $H < 1/2$ or $H > 1/2$), and differs only by the integration limits. 
 Therefore, our results show that these three kinds of anomalous diffusions 
are indeed the members of the same family.

\section*{Acknowledgments}

The authors acknowledge fruitful and instructive discussions with Baruch Meerson, Karol Penson and Horacio Wio.

\begin{appendix}

\section{Sub-diffusive one-sided Mandelbrot-van Ness fBm}\label{secA1}

{\bf Representation in eq. \eqref{action1ssub}.} The integral in eq. \eqref{BQ} can be performed explicitly to give
\begin{align}
\label{Bqq}
K(t_1,t_2)  &=  \frac{{\rm ctg}(\pi H)}{\pi H (1 - 2H) B(1/2-H,2H)} \nonumber\\
&\times \frac{d^2}{dt_1 dt_2}  \frac{(t_1 t_2)^{1/2 - H}}{\left(t_1 + t_2\right)} \,_2F_1\left(\frac{1}{2},1; \frac{3}{2} - H; \frac{4 t_1 t_2}{(t_1 + t_2)^2}\right) \,.
\end{align}
The Gauss hypergeometric function in the right-hand-side  of eq. \eqref{Bqq} can be expressed through
the incomplete beta function, which yields, upon a substitution into eq. \eqref{action} and 
the integration by parts, the representation \eqref{action1ssub}. \\
{\bf Limit $H \to 1/2$.} To access the behaviour in the limit $H \to 1/2$, (in which the expressions  \eqref{action1ssub} and \eqref{main1a} should evidently converge to the expression \eqref{action0}), it is convenient to formally rewrite the Gauss hypergeometric function in eq. \eqref{Bqq}  as a linear combination of two hypergeometric functions. In doing so, we get
\begin{align}
\,_2F_1\left(\frac{1}{2},1; \frac{3}{2} - H; z\right) = \frac{\Gamma(H) \Gamma(3/2-H)}{\sqrt{\pi} (1-z)^{H} z^{1/2-H}} - \frac{(1/2-H)}{H} \,_2F_1\left(\frac{1}{2},1; 1+ H; 1-z\right) \,, 
\end{align} 
 such that we find a representation of the kernel $K(t_1,t_2)$ in which a singular (in the limit $t_1 \to t_2$)  part enters additively to a regular at this point contribution:
\begin{align}
\label{dom}
K(t_1,t_2)  &=  \frac{{\rm ctg}(\pi H)}{2 \pi H} \frac{d^2}{dt_1 dt_2}  \frac{1}{|t_1 - t_2|^{2H}}  \nonumber\\
&- \frac{{\rm ctg}(\pi H)}{2 \pi H^2 B(1/2-H,2H)} \frac{d^2}{dt_1 dt_2}  \frac{(t_1 t_2)^{1/2-H}}{(t_1 + t_2)} \,_2F_1\left(\frac{1}{2},1; 1+ H; \left(\frac{t_1 - t_2}{t_1 + t_2}\right)^2\right) \,.
\end{align}
In the limit $H \to 1/2$, the numerical factor in front of the second term in eq. \eqref{dom} vanishes, 
because
\begin{equation}
\frac{{\rm ctg}(\pi H)}{2 \pi H^2 B(1/2-H,2H)} = 2 \left(1/2- H\right)^2 + O\left(\left(1/2-H\right)^3\right) \,,
\end{equation}
and hence, the second term becomes equal to zero. In turn, to determine the limiting form of the first term in eq. \eqref{dom} we use the identity
\begin{equation}
\label{ident1}
\frac{1}{|t_1-t_2|^{2H}} = \frac{1}{(1 - 2 H)} \frac{d}{dt_1} {\rm sign}(t_1 - t_2) |t_1 - t_2|^{1 - 2H} \,,
\end{equation}
such that the first term in eq. \eqref{dom} can be formally rewritten as
\begin{equation}
\label{ident2}
 \frac{{\rm ctg}(\pi H)}{2 \pi H} \frac{d^2}{dt_1 dt_2}  \frac{1}{|t_1 - t_2|^{2H}}  = \frac{{\rm ctg}(\pi H)}{2 \pi H (1 - 2 H)}  \frac{d^3}{dt_1^2 dt_2} {\rm sign}(t_1 - t_2) |t_1 - t_2|^{1 - 2H} \,.
\end{equation}
The latter form implies that
\begin{equation}
\label{ident3}
\lim_{H \to 1/2}  \frac{{\rm ctg}(\pi H)}{2 \pi H} \frac{d^2}{dt_1 dt_2}  \frac{1}{|t_1 - t_2|^{2H}} = \frac{1}{2} \frac{d^3}{dt_1^2 dt_2}  {\rm sign}(t_1 - t_2) = \frac{d^2}{dt_1 dt_2}  \delta(t_1 - t_2)
\end{equation}
Therefore, we automatically recover the expression \eqref{action0}.

\section{Sub-diffusive two-sided Mandelbrot-van Ness fBm}\label{secA2}

{\bf Integral equation for $K(t_1,t_2)$.} Likewise in the one-sided case, we differentiate first both sides of eq. \eqref{int} with respect to variable $t_3$ to get
\begin{align}
\label{A}
\int_{-\infty}^{\infty} dt_1 \, K(t_1,t_2) \, \frac{d}{dt_3} {\rm Cov}\left(t_1,t_3\right) = \frac{d}{dt_3} \delta\left(t_2 - t_3\right) = - \frac{d}{dt_2} \delta\left(t_2 - t_3\right) \,.
\end{align}
For the two-sided fBm, the derivative of the covariance function is given by
\begin{align}
\label{der}
\frac{d}{dt_3} {\rm Cov}\left(t_1,t_3\right) &=  
H \left(\frac{{{\rm sign}(t_3)}}{|t_3|^{1 - 2H}} - \frac{{{\rm sign}(t_3 - t_1)}}{|t_3 - t_1|^{1 - 2H}}\right) \,.
\end{align}
Inserting the latter expression into eq. \eqref{A}, and rearranging the resulting equation, we arrive at the following integral equation with a  Feller potential in the left-hand-side :
\begin{align}
\label{B}
\int_{-\infty}^{\infty} dt_1 \, K(t_1,t_2) \, \frac{{{\rm sign}(t_3 - t_1)}}{|t_3 - t_1|^{1 - 2H}} = \frac{1}{H} \frac{d}{dt_2} \delta\left(t_2 - t_3\right) + \frac{{{\rm sign}(t_3)}}{|t_3|^{1 - 2H}} \left(\int_{-\infty}^{\infty} du \, K(u,t_2)\right) \,.
\end{align}
Note first that this integral equation  is not closed with respect to the kernel $K(t_1,t_2)$ -- it  
contains  in its right-hand-side this function integrated over $t_1$, which is a function of $t_2$. We thus have to solve eq. \eqref{B} supposing that this integrated kernel is some known function of $t_2$, 
and then to define it self-consistently. 
Note, as well, that because of the differentiation of eq. \eqref{int}  with respect to 
 $t_3$, we have evidently lost  some additional $t_3$-independent terms in the right-hand-side of eq. \eqref{B}, which renders the solution of this equation to be not a symmetric function of $t_1$ and $t_2$, while in reality it should be a symmetric one. 
 Hence, some additional arguments will be needed to restore this symmetry. \\
{\bf Solution of eq. \eqref{B}.} 
The solution of eq. \eqref{B} is found by 
applying the inverse Feller transform  (see, e.g.,  \cite{kilbas}) and reads :
\begin{align}
\label{solution}
K(t_1,t_2) &= \frac{{\rm ctg}(\pi H)}{2 \pi} \frac{d}{dt_1} \int^{\infty}_{-\infty} \frac{dt}{|t_1 - t|^{2H}} \Bigg( \frac{1}{H} \frac{d}{dt_2} \delta\left(t_2 - t\right) + \frac{{{\rm sign}(t)}}{|t|^{1 - 2H}} \left(\int_{-\infty}^{\infty} du \, K(u,t_2)\right) \Bigg) \nonumber\\
&= \frac{{\rm ctg}(\pi H)}{2 \pi H} \frac{d^2}{dt_1 dt_2} \frac{1}{|t_1 - t_2|^{2H}} + \frac{{\rm ctg}(\pi H)}{2 \pi} \left(\int_{-\infty}^{\infty} du \, K(u,t_2) \right)  \frac{d}{dt_1} L(t_1)  \,,
\end{align}
where
\begin{align}
 L(t_1) = \int^{\infty}_{-\infty} \frac{{\rm sign}(t) dt}{|t|^{1 - 2H} |t_1 - t|^{2H}} = \pi \, {\rm tg}(\pi H) \, {\rm sign}(t_1)  \,,
\end{align} 
such that
\begin{align}
\frac{d}{dt_1} L(t_1) = 2 \pi \,{\rm tg}(\pi H) \delta(t_1) \,.
\end{align}
This means that the $t_1$-dependence of the second term in the right-hand-side of eq. \eqref{solution} is merely the delta-function of $t_1$, and this singular term equals zero everywhere except for $t_1 = 0$.
We note that such a local term in the solution stems from the cusp-like behaviour of the covariance function in eq. \eqref{22}, when either of the time variables changes its sign, and hence, the full solution should have the following structure : 
\begin{align}
\label{sol1}
K(t_1,t_2)  &=  \frac{{\rm ctg}(\pi H)}{2 \pi H} \frac{d^2}{dt_1 dt_2} \frac{1}{|t_1 - t_2|^{2H}} \nonumber\\
&+ \left(\int_{-\infty}^{\infty} dv \, K(t_1,v) \right)  \delta(t_2) + \left(\int_{-\infty}^{\infty} du \, K(u,t_2) \right)  \delta(t_1) \,.
\end{align}
Further on, on intuitive grounds,
 we conjecture that the kernel $K(t_1,t_2)$ integrated over either of the variables obeys
 \begin{align}
 \label{sol2}
 \int_{-\infty}^{\infty} dv \, K(t_1,v) = - \frac{1}{H} \frac{d}{dt_1} \frac{t_1}{|t_1|^{2H}} \delta(t_1) \,, \,\,\, \int_{-\infty}^{\infty} du \, K(u,t_2)  =  - \frac{1}{H} \frac{d}{dt_2} \frac{t_2}{|t_2|^{2H}} \delta(t_2) \,,
 \end{align}
 such that, explicitly,
 \begin{align}
\label{sol3}
K(t_1,t_2)  &=  \frac{{\rm ctg}(\pi H)}{2 \pi H} \frac{d^2}{dt_1 dt_2} \frac{1}{|t_1 - t_2|^{2H}} \nonumber\\
&- \frac{\delta(t_2)}{H} \frac{d}{dt_1} \frac{t_1}{|t_1|^{2H}} \delta(t_1) -  \frac{\delta(t_1)}{H}  \frac{d}{dt_2} \frac{t_2}{|t_2|^{2H}} \delta(t_2) \,.
\end{align} 
It is straightforward to verify that the expression \eqref{sol3} becomes an identity, if one integrates both sides over $t_1$ or $t_2$.  
 The question now is if the expression \eqref{sol3} 
 is the full solution of the integral equation \eqref{int}. 
Below we show that it is indeed the case. 

We first check directly whether the first "regular" term in eq. \eqref{sol1} fulfills the integral equation \eqref{int}.
Using the integral identity
\begin{align}
\frac{|t|^{2H}}{2} = \frac{\sin(\pi H) \Gamma(2H+1)}{2 \pi}  \int^{\infty}_{-\infty}  \frac{|\lambda|^{1-2H}}{\lambda^2} \left(1 - e^{i \lambda t}\right) d\lambda  \,, 
\end{align}
the covariance function of the two-sided fBm, eq. \eqref{22}, 
can be cast into a convenient factorised form (see e.g., \cite{yaglom}) :
\begin{align}
\label{yaglom}
{\rm Cov}(t_1,t_3) =  \frac{\sin(\pi H) \Gamma(2H+1)}{2 \pi}  \int^{\infty}_{-\infty}  \frac{|\lambda|^{1-2H}}{\lambda^2} \left(1 - e^{i \lambda t_1}\right) \left(1 - e^{- i \lambda t_3}\right) d\lambda \,.
\end{align} 
Concurrently, the function in the first term in eq. \eqref{sol3} can be written formally as the Fourier-transform :
\begin{align}
\label{kernel}
&\frac{{\rm ctg}(\pi H)}{2 \pi H} \frac{1}{|t_1 - t_2|^{2H}}  = \frac{1}{2 \pi \sin(\pi H) \Gamma(2H+1)}  \int^{\infty}_{-\infty} |p|^{2H - 1} dp \, e^{i p (t_1 - t_2)}  \,.
\end{align}
Then, we have
\begin{align}
& \frac{{\rm ctg}(\pi H)}{2 \pi H}  \int^{\infty}_{-\infty} dt_1 \, {\rm Cov}(t_1,t_3)  \frac{d^2}{dt_1 dt_2} \frac{1}{|t_1 - t_2|^{2H}}  = \frac{1}{(2 \pi)^2} \int^{\infty}_{-\infty} |p|^{2H + 1} dp \, e^{- i p  t_2} \nonumber\\
& \times \int^{\infty}_{-\infty}  \frac{|\lambda|^{1-2H}}{\lambda^2}  \left(1 - e^{- i \lambda t_3}\right) d\lambda \int^{\infty}_{-\infty} dt_1 \, e^{i p t_1} \, \left(1 - e^{i \lambda t_1}\right)
\nonumber\\
&=\frac{1}{2 \pi} \int^{\infty}_{-\infty} |p|^{2H + 1} dp \, e^{- i p  t_2}  \int^{\infty}_{-\infty}  \frac{|\lambda|^{1-2H}}{\lambda^2}  \left(1 - e^{- i \lambda t_3}\right) \Big(\delta(p) - \delta(p+\lambda)\Big) d\lambda \nonumber\\
&= \delta(t_3 - t_2) - \delta(t_2) \,.
\end{align}
Therefore, the regular term alone is not the full solution of eq. \eqref{int} and generates an uncompensated delta-function $\delta(t_2)$, which appears with sign $"-"$.

We check next the contribution of the second and the third terms in eq. \eqref{sol3}. The third term, proportional to $\delta(t_1)$ trivially gives a zero contribution, because ${\rm Cov}(t_1 = 0,t_3) = 0$. The contribution of the second term is
\begin{align}
&- \frac{\delta(t_2)}{H}  \int^{\infty}_{-\infty} dt_1 {\rm Cov}(t_1,t_3) \frac{d}{dt_1} \frac{t_1}{|t_1|^{2H}} \delta(t_1) = - \frac{(1-2H) \delta(t_2)}{H} \int^{\infty}_{-\infty} dt_1 \frac{{\rm Cov}(t_1,t_3)}{|t_1|^{2H}} \delta(t_1) \nonumber\\&- \frac{\delta(t_2)}{H} \int^{\infty}_{-\infty} dt_1 \frac{t_1 {\rm Cov}(t_1,t_3)}{|t_1|^{2H}}  \frac{d}{dt_1} \delta(t_1) = - \frac{(1-2H) \delta(t_2)}{2 H} + \nonumber\\&+ \frac{\delta(t_2)}{H} \int^{\infty}_{-\infty} dt_1 \left(\frac{d}{dt_1} \frac{t_1 {\rm Cov}(t_1,t_3)}{|t_1|^{2H}} \right) \delta(t_1) = - \frac{(1-2H) \delta(t_2)}{2 H} + \frac{\delta(t_2)}{2 H} = \delta(t_2) \,,
\end{align}
which thus removes  $"- \delta(t_2)"$ produced by the first term. As a consequence,  the expression in eq. \eqref{sol3}  is indeed the full solution of the integral equation \eqref{int}.\\
{\bf Representation in  eq. \eqref{action2ssub}.} 
Consider next the action \eqref{action} with the kernel \eqref{sol3}. Inserting the expression  \eqref{sol3} into eq. \eqref{action} we have
\begin{align}
\label{L1}
S[x(t)] &= \frac{{\rm ctg}(\pi H)}{4 \pi H} \int_{-\infty}^{\infty} dt_1 \, x(t_1) \int_{-\infty}^{\infty} dt_2 \, x(t_2) \, \frac{d^2}{dt_1 dt_2} \frac{1}{|t_1-t_2|^{2H}} \nonumber\\
&-\frac{x(t=0)}{H} \left(\int^{\infty}_{-\infty} dt \, x(t) \, \frac{d}{dt} \frac{t}{|t|^{2H}} \delta(t)\right)  \,.
\end{align}
Since $x(t=0) \equiv 0$ by definition, unless the integral in the brackets diverges, which may happen for such trajectories whose time-derivative diverges at the origin very steeply, i.e.,
\begin{align}
\lim_{t \to 0^+, t\to 0^-} \frac{t}{|t|^{2H}} \frac{dx(t)}{dt} \to \pm \infty \,,
\end{align}
the term in the second line in eq. \eqref{L1}  is therefore equal to zero. Performing  the integration by parts  in the remaining term, we arrive at eq. \eqref{action2ssub} derived previously in \cite{we} using a slightly different approach. Convergence of the action in eq. \eqref{action2ssub} to the standard result in eq. \eqref{action0} in the limit $H \to 1/2$ follows trivially from our eqs. \eqref{ident2} and \eqref{ident3}.

\section{Super-diffusive one-sided Mandelbrot-van Ness fBm}\label{secA3}

{\bf Solution of the integral equation \eqref{int}.} Inserting the  identity \eqref{rep} into eq. \eqref{int} and integrating the left-hand-side of the resulting equation by parts, we have
\begin{equation}
\begin{split}
\label{b}
&\left.  \left(\int^{t_1}_0 d\tau_1 K(\tau_1,t_3)\right) \left(\int^{t_1}_0 \int^{t_2}_0 \frac{du dv}{|u-v|^{2-2H}}\right)\right|^{t_1 = \infty}_{t_1=0}\\&- \int^{\infty}_0 dt_1 \left(\int^{t_1}_0 d\tau_1 \, K(\tau_1,t_3)\right) \int^{t_2}_0 \frac{dv}{|t_1 -v|^{2- 2H}} = \frac{1}{H (2H - 1)} \delta(t_2 - t_3) \,.
\end{split}
\end{equation}
Evidently, the first term in the left-hand-side of the latter equation vanishes for $t_1=0$. Below we show that it also vanishes when $t_1 \to \infty$, such that the first term is equal to zero. Consequently, eq. \eqref{b}  reduces to
\begin{equation}
\label{bb}
- \int^{\infty}_0 dt_1 \left(\int^{t_1}_0 d\tau_1 K(\tau_1,t_3)\right) \int^{t_2}_0 \frac{dv}{|t_1 -v|^{2- 2H}} = \frac{1}{H (2H - 1)} \delta(t_2 - t_3) \,.
\end{equation}
We integrate both sides of eq. \eqref{bb} over $t_3$, from zero to $t_3$,  which gives, taking into account that
\begin{equation}
\int^{t_3}_0 d\tau \, \delta(t_2 - \tau) 
= \theta(t_3 - t_2)
\end{equation}
the following integral equation :
\begin{equation}
\label{b3}
- \int^{\infty}_0 dt_1 \left(\int^{t_1}_0 d\tau_1 \int^{t_3}_0 d\tau_3 \, K(\tau_1,\tau_3)\right) \int^{t_2}_0 \frac{dv}{|t_1 -v|^{2- 2H}} = \frac{1}{H (2H - 1)} \theta(t_3 - t_2) \,.
\end{equation}
Differentiating both sides of eq. \eqref{b3} over $t_2$ and 
introducing an auxiliary function  $q(t_1,t_3) $ (see eq. \eqref{quq}), we arrive at our eq. \eqref{int3}.\\
{\bf Representation in eq. \eqref{main2a}.} The formal solution of the latter integral equation is given by (see \cite{lundgren})
\begin{equation}
\begin{split}
\label{d}
q(t_1,t_3) &= - \frac{1}{2 H \sin(\pi H) \Gamma(2H) \Gamma^2(3/2-H)} \frac{d}{dt_1} \int^{t_1}_0 \frac{d\tau}{(t_1 - \tau)^{H-1/2}} \frac{d}{d\tau} \int^{\infty}_{\tau} \frac{d\gamma \, \delta(t_3 - \gamma)}{(\gamma - \tau)^{H-1/2}} \,.
\end{split}
\end{equation}
Performing the inner integral over $d\gamma$, eq. \eqref{d} can be somewhat simplified to give
\begin{equation}
\begin{split}
\label{dd}
q(t_1,t_3) &= - \frac{1}{2 H \sin(\pi H) \Gamma(2H) \Gamma^2(3/2-H)} \frac{d}{dt_1} \int^{t_1}_0 \frac{d\tau}{(t_1 - \tau)^{H-1/2}} \frac{d}{d\tau}  \frac{\theta(t_3-\tau)}{(t_3 - \tau)^{H-1/2}} \\
&= \frac{1}{2 H \sin(\pi H) \Gamma(2H) \Gamma^2(3/2-H)} \frac{d^2}{dt_1 dt_3} \int^{{\rm min}(t_1,t_3)}_0 \frac{d\tau}{[(t_1 - \tau)  (t_3 - \tau)]^{H-1/2} }  \,,
\end{split}
\end{equation}
where the expression in the last line is, in fact, our eq. \eqref{qus}, which yields eventually the representation  in eq. \eqref{main2a}. \\
{\bf Representation in eq. \eqref{action1ssuper}.} Explicit form of action in eq. \eqref{action1ssuper} is determined as follows:
The integral in the last line in eq. \eqref{dd} can be performed exactly
\begin{equation}
\begin{split}
\label{s}
\int^{{\rm min}(t_1,t_3)}_0 \frac{d\tau}{[(t_1 - \tau)   (t_3 - \tau)]^{H-1/2} } &= |t_1 - t_3|^{2-2H} B_{z'}(3/2-H,2H-2) \\&
z'=\frac{{\rm min}(t_1,t_3)}{{\rm max}(t_1,t_3)} \,,
 \end{split}
\end{equation} 
where $B_{z'}(a,b)$ is the incomplete beta-function. Consequently, $q(t_1,t_3)$ is given explicitly by
\begin{equation}
\label{mkn}
q(t_1,t_3) = \frac{1}{2 H \sin(\pi H) \Gamma(2H) \Gamma^2(3/2-H)} \frac{d^2}{dt_1 dt_3}  |t_1 - t_3|^{2-2H} B_{z'}(3/2-H,2H-2)  \,.
\end{equation}
We rewrite eq. \eqref{mkn} in terms of the regularised incomplete beta-function $I_z(a,b)$,
\begin{equation}
B_{z'}(3/2-H,2H-2) = \frac{\Gamma(3/2-H) \Gamma(2H-2)}{\Gamma(H - 1/2)}       I_z(3/2-H,2H-2) \,,
\end{equation}
to get  the action in eq. \eqref{action1ssuper}.\\
{\bf First  term in eq. \eqref{b}.} The term in the first line in eq. \eqref{b} evidently vanishes at $t_1 = 0$. We concentrate 
on its behaviour in the limit $t_1 \to \infty$ (for $t_2 $ and $t_3$ kept fixed). In this limit, $q(t_1,t_3)$ in eq. \eqref{mkn} behaves as
\begin{equation}
q(t_1,t_3) \simeq \frac{1}{t_3^{H-1/2} t_1^{H+1/2}} \,,  \quad t_1 \to \infty \,.
\end{equation}
Consequently, the first integral in the first term in eq. \eqref{b} follows, in virtue of eq. \eqref{quq}, 
\begin{equation}
\int^{t_1}_0 d\tau_1 K(\tau_1,t_3) = \frac{d}{dt_3} q(t_1,t_3) \simeq \frac{1}{t_3^{H+1/2} t_1^{H+1/2}} \,,  \quad t_1 \to \infty \,.
\end{equation}
In turn, the two-fold integral in the first term in eq. \eqref{b} is simply the covariance function of the one-sided Mandelbrot-van Ness fBm. Hence, we have
\begin{equation}
\int^{t_1}_0 \int^{t_2}_0 \frac{du dv}{|u-v|^{2-2H}} \simeq t_2 t_1^{2H - 1} \,, \quad t_1 \to \infty \,,
\end{equation}
such the first term vanishes in the limit $t_1 \to \infty$ in proportion to $(t_2/t_3^{H+1/2}) t_1^{-3/2+H}$.\\
{\bf Limit $H \to 1/2$.} We can set $H = 1/2$ directly in eq. \eqref{d}, which gives
\begin{equation}
\begin{split}
\label{d4}
q(t_1,t_3) &= -  \frac{d}{dt_1} \int^{t_1}_0 d\tau \frac{d}{d\tau} \int^{\infty}_{\tau} d\gamma \, \delta(t_3 - \gamma)\\
&=  -  \frac{d}{dt_1} \int^{t_1}_0 d\tau \frac{d}{d\tau} \theta(t_3 - \tau)  = \frac{d}{dt_1} \int^{t_1}_0 d\tau \, \delta(t_3 - \tau)\\
& =  \frac{d}{dt_1} \theta(t_1 - t_3) = \delta(t_1 -  t_3) \,.
\end{split}
\end{equation}
This  proves that the action converges to the one defined in eq. \eqref{action0} in the limit $H \to 1/2$. 

\section{Super-diffusive two-sided Mandelbrot-van Ness fBm}\label{secA4}

{\bf Equation for $u(t_1,t_2)$ and its solution.} Integrating eq. \eqref{equ} over $t_2$ from minus infinity till $t_2$, we have
\begin{equation}
\label{equ1}
\int^{\infty}_{-\infty} \frac{dt_1 \, u(t_1,t_2)}{|t_1 - t_3|^{2-2H}}  = \frac{1}{H (2H  - 1)} \int_{-\infty}^{t_2} \delta(\tau_2 - t_3) d\tau_2 = \frac{1}{H (2H  - 1)} \theta(t_2 - t_3) \,.
\end{equation}
Next, we take advantage of the integral identity, (which is, in fact, the same as the one in eq. \eqref{zuk})
\begin{equation}
\frac{1}{|t_1 - t_3|^{2-2H}} = \frac{\Gamma(3/2 - H)}{\Gamma(H-1/2) \Gamma(2 - 2 H)} \int^{\infty}_{-\infty} dt \frac{\theta(t-t_1) \theta(t-t_3)}{[(t-t_1) (t-t_3)]^{3/2-H}}  \,, 
\end{equation}
which permits to formally rewrite eq. \eqref{equ1} as
\begin{equation}
\label{qs}
\int^{\infty}_{t_3} \frac{dt \, \phi(t,t_2)}{(t-t_3)^{3/2-H}}  = - \frac{\Gamma(H-1/2) \Gamma(1-2H)}{H \Gamma(3/2-H)} \theta(t_2 - t_3) \,.
\end{equation}
with
\begin{equation}
\label{phi}
\phi(t,t_2) = \int^{t}_{-\infty} \frac{dt_1 u(t_1,t_2)}{(t-t_1)^{3/2-H}} \,.
\end{equation}
Multiplying both sides (from the left) of eq. \eqref{qs} by the integral operator
\begin{equation}
\int^{\infty}_z \frac{dt_3}{(t_3 - z)^{H-1/2}}  \times \,,
\end{equation}
i.e., representing eq. \eqref{qs} as
\begin{equation}
\begin{split}
\int^{\infty}_z \frac{dt_3}{(t_3 - z)^{H-1/2}} \int^{\infty}_{t_3} \frac{dt \, \phi(t,t_2)}{(t-t_3)^{3/2-H}}  &= - \frac{\Gamma(H-1/2) \Gamma(1-2H)}{H \Gamma(3/2-H)} \\&\times \int^{\infty}_z \frac{dt_3}{(t_3 - z)^{H-1/2}}  \theta(t_2 - t_3)
\end{split}
\end{equation}
and interchanging the order of integrations in the left-hand-side of the latter equation via the Dirichlet formula
\begin{equation}
\label{dir}
\int^b_a dx \int^x_a f(x,y) dy =\int^b_a dy \int^b_y f(x,y) dy \,,
\end{equation}
we obtain
\begin{equation}
\begin{split}
\label{equ3}
\int^{\infty}_z dt \, \phi(t,t_2)  \int^{t}_z \frac{dt_3}{(t-t_3)^{3/2-H}(t_3 - z)^{H-1/2}}  &= - \frac{\Gamma(H-1/2) \Gamma(1-2H)}{H \Gamma(3/2-H)}\\&\times \int^{\infty}_z \frac{dt_3}{(t_3 - z)^{H-1/2}}  \theta(t_2 - t_3) \,.
\end{split}
\end{equation}
Further on, one observes that
\begin{equation}
\int^{t}_z \frac{dt_3}{(t-t_3)^{3/2-H}(t_3 - z)^{H-1/2}} = - \frac{\pi}{\cos(\pi H)}
\end{equation}
and hence,
eq. \eqref{equ3} becomes
\begin{equation}
\int^{\infty}_z dt \, \phi(t,t_2) = \frac{\cos(\pi H) \Gamma(H-1/2) \Gamma(1-2H)}{\pi H \Gamma(3/2-H)} \int^{\infty}_z \frac{dt_3}{(t_3 - z)^{H-1/2}}  \theta(t_2 - t_3) \,.
\end{equation}
Differentiating both sides of the latter equation with respect to $z$, we obtain
\begin{equation}
\begin{split}
 \phi(z,t_2) &= - \frac{\cos(\pi H) \Gamma(H-1/2) \Gamma(1-2H)}{\pi H \Gamma(3/2-H)} \frac{d}{dz} \int^{\infty}_z \frac{dt_3}{(t_3 - z)^{H-1/2}}  \theta(t_2 - t_3) \\
 &=\frac{\cos(\pi H) \Gamma(H-1/2) \Gamma(1-2H)}{\pi H \Gamma(3/2-H)} \frac{\theta(t_2-z)}{(t_2 - z)^{H-1/2}}
 \end{split}
\end{equation}
or, in virtue of eq. \eqref{phi}, 
\begin{equation}
\int^{z}_{-\infty} \frac{dt_1 u(t_1,t_2)}{(z-t_1)^{3/2-H}}  = \frac{\cos(\pi H) \Gamma(H-1/2) \Gamma(1-2H)}{\pi H \Gamma(3/2-H)} \frac{\theta(t_2-z)}{(t_2 - z)^{H-1/2}}
\end{equation}
The latter integral equation can be solved again by applying from the left to both sides an appropriate fractional integral operator,
\begin{equation}
\begin{split}
\int^{\xi}_{-\infty} \frac{dz}{(\xi - z)^{H-1/2}} \int^{z}_{-\infty} \frac{dt_1 u(t_1,t_2)}{(z-t_1)^{3/2-H}}  &= \frac{\cos(\pi H) \Gamma(H-1/2) \Gamma(1-2H)}{\pi H \Gamma(3/2-H)} \\&\times \int^{\xi}_{-\infty} \frac{dz}{(\xi - z)^{H-1/2}} \frac{\theta(t_2-z)}{(t_2 - z)^{H-1/2}}
\end{split}
\end{equation}
Interchanging  the order of integration in the left-hand-side using the Dirichlet formula \eqref{dir} and performing the inner integral we find
\begin{equation}
\label{sm}
\int^{\xi}_{-\infty} u(t_1,t_2) dt_1 =  - \frac{{\rm ctg}(\pi H) \Gamma(H-1/2)}{2 \pi H \Gamma(3/2-H) \Gamma(2 H)} \int^{\xi}_{-\infty} \frac{dz}{(\xi - z)^{H-1/2}} \frac{\theta(t_2-z)}{(t_2 - z)^{H-1/2}} \,,
\end{equation}
from which our eq. \eqref{equ5} follows by a mere differentiation with respect to variable $\xi$.\\
{\bf Explicit form in eq. \eqref{action2ssuper}.} In case when $t_1 > t_2$, the right-hand-side of eq. \eqref{equ5} can be calculated very straightforwardly - we first differentiate the integrand over $t_1$ and then perform the integral. This gives
\begin{equation}
u(t_1,t_2)  =  \frac{{\rm ctg}(\pi H)}{2 \pi H (2H - 1)} \frac{1}{(t_1 - t_2)^{2H - 1}}  =  \frac{{\rm ctg}(\pi H)}{4 \pi H (1-H) (2H - 1)} \frac{d}{dt_1} (t_1 - t_2)^{2 - 2 H}\,.
\end{equation}
In case when $t_2 > t_1$ we first change the integration variable for  $u = t_1 - z$, differentiate the integrand with respect to $t_1$ and then perform the integral. This yields
\begin{equation}
u(t_1,t_2)  = -  \frac{{\rm ctg}(\pi H)}{2 \pi H (2H - 1)} \frac{1}{(t_2 - t_1)^{2H - 1}} = \frac{{\rm ctg}(\pi H)}{4 \pi H (1-H) (2H - 1)} \frac{d}{dt_1} (t_2 - t_1)^{2 - 2 H}\,.
\end{equation}
Consequently, for any relation between $t_1$ and $t_2$ we have
\begin{equation}
\begin{split}
u(t_1,t_2)  &= \frac{{\rm ctg}(\pi H)}{2 \pi H (2H - 1)} \frac{{\rm sign}(t_1 - t_2)}{|t_1 - t_2|^{2H - 1}} \\&= 
\frac{{\rm ctg}(\pi H)}{4 \pi H (1-H) (2H - 1)} \frac{d}{dt_1} |t_1 - t_2|^{2 - 2 H}\,.
\end{split}
\end{equation}
Inserting this expression into the action in the first line in eq. \eqref{lab} and integrating by parts with respect to the variable $t_1$, we get the explicit form of the action in eq. \eqref{action2ssuper} \cite{we}.\\
{\bf Limit $H \to 1/2$.} The form of the action in eq. \eqref{main2b}
in the limit $H \to 1/2$ can be conveniently accessed by setting $H = 1/2$ directly  in eq. \eqref{lab}. 
Noticing that
\begin{equation}
\lim_{H\to 1/2} \frac{1}{4 H \sin(\pi H) \Gamma(2H)} = \frac{1}{2} \,,
\end{equation}
and 
\begin{equation}
 \int^{\infty}_t \frac{d^2 x(t_2)}{dt_2^2} dt_2 = - \frac{dx(t)}{dt} \,, \quad 
 \int^{\infty}_{-\infty} dt_1 \frac{d x(t_1)}{dt_1} \frac{d}{dt_1} \theta(t_1- t) =  \frac{dx(t)}{dt} \,,
\end{equation}
we recover the standard result for Brownian motion in eq. \eqref{action0}.

\end{appendix}

\end{document}